\documentclass[a4paper,10pt]{article}

\usepackage{graphicx}
 \usepackage{setspace}
\title{Quantum tasks using six qubit cluster states}
\author{Jayakrishnan V Menon$^{a}$, Naveen Paul$^{b}$, Siddharth Karumanchi$^{c}$,\\ Sreraman Muralidharan$^{d}$ and Prasanta K.Panigrahi$^{e}$ \\\\
$^{a}$ Christ University, Hosur Road, Bangalore- 560029, India\\
$^{b}$ M.N.M. Jain Engineering College, Chennai- 600096, India\\
$^{c}$ Birla Institute of Technology and Science, Pilani, India \\
$^{d}$ Loyola College, Nungambakkam, Chennai - 600 034, India\\
$^{e}$ Indian Institute of Science Education and Research-Kolkata, Mohanpur Campus, \\BCKV Campus Main Office, Mohanpur - 741252, India and\\ Physical Research Laboratory, Navrangpura, Ahmedabad - 380 009, India\\\\
}
\begin{document}
\maketitle

\title{}
\author{}

\maketitle

\begin{abstract}

The usefulness of the recent experimentally realized six photon cluster state by C. Y. Lu \it {et al.} \normalfont(2007 \it{Nature} \normalfont \bf{3} \normalfont {91}), is investigated for quantum communication protocols like teleportation, quantum information splitting (QIS), remote state preparation and dense coding. We show that the present state can be used for the teleportation of an arbitrary two qubit state deterministically. Later we devise two distinct protocols for the QIS of an arbitrary two qubit state among two parties and systematically
compare their relative merits in terms of classical communication and security.
Sixteen orthogonal measurement basis on the cluster state is constructed, which will lock an arbitrary two qubit state
among two parties. The usefulness of the state for dense coding is investigated and it is shown that one can send five classical bits by sending only three qubits using this state as a shared
entangled resource. We finally show that this state can also be utilised in the remote state preparation of an arbitrary two qubit state. 

\end{abstract}

\section{Introduction}

Entanglement helps in carrying out several quantum tasks like teleportation \cite{Bennett}, secret sharing \cite{Hillery, Gott}, dense coding \cite{Wiesner} and one-way quantum computation \cite{one}. Multi-partite entangled states, arising from different physical systems, have been used to achieve these purposes. The way in which a given state is entangled plays a major role in deciding its suitability to perform a certain quantum task. Hence, not all entangled states can be used for the desired purposes. For instance, while a three qubit $GHZ$ state,
can be used for the teleportation of an arbitrary single qubit state $|\psi_1\rangle = (\alpha|0\rangle + \beta|1\rangle)$, where
$\alpha$, $\beta$ $\in$ C and $|\alpha|^2+|\beta|^2$=1,  a symmetric $W$ state  cannot be used for the same \cite{wstate}. In the case of four qubits, entanglement has been classified into
nine categories under LOCC \cite{class}. Only one of these nine classes can be used for the teleportation of an arbitrary two qubit state $|\psi_{2}\rangle=\alpha|00\rangle + \mu|10\rangle + \gamma|01\rangle + \beta|11\rangle$, where $|\alpha|^2+|\mu|^2+|\gamma|^2+|\beta|^2$=1 and $\alpha,$ $\mu,$  $\gamma,$ $\beta \in C$. Recently
\cite{Brown, Borras}, new multipartite entangled channels have been constructed based on numerical optimisation schemes and their efficiency have been checked for various quantum protocols. For instance, it has been shown \cite{Sreramanb, Sayan} that the five and six qubit states respectively, introduced by Brown \it {et al. }\normalfont \cite{Brown}, \normalfont and Borras \it {et al.} \normalfont\cite{Borras}, through extensive search procedures, can be used for various quantum protocols. Though a number of quantum protocols have been explicated theoretically, only a few of them are experimentally realizable. Hence, deriving quantum protocols using multipartite entangled states, which have been experimentally realized, is of immense interest in quantum information theory. It is worth mentioning that in the experimental scenario, entanglement has been achieved between 
six qubits \cite{expt}.

In recent years, special types of entangled states known as the graph states \cite{Graph} have attracted much attention owing to their promising usefulness in quantum information theory. In general a graph state is associated with a graph, in which each vertex is a Hadamard state and each edge represents a control phase shift interaction. The multipartite entanglement revealed by these states depend on the geometry of the underlining graph. The two important types of multipartite graph states are the GHZ states :
\begin{equation}
|GHZ\rangle_N = \frac{1}{\sqrt{2}} (|0\rangle^{\otimes N} + |1\rangle^{\otimes N}),
\end{equation}
and cluster states \cite{one}:
\begin{equation}
|C\rangle_N = \frac{1}{2^{N/2}} \otimes_{a=1}^{N} (|0\rangle_a \sigma_z ^{a+1} + |1\rangle_a).
\end{equation}
In general, the $N$-qubit cluster state belongs to the class of Bell and the GHZ state for $N = 2$, $N =3 $ respectively and it exhibits different entanglement properties from the GHZ states for $N \geq 4$ under LOCC. These states have been proved to be useful for one way quantum computation \cite{Robert} and for quantum error correction \cite{Sch}. They also show a strong violation of reality and are robust against decoherence \cite{local}. One way quantum computing has been experimentally demonstrated using the cluster states \cite{expt}. Interestingly, while there exists one entangled bit between any two subsystems in a $N$-qubit GHZ state, there exists two entangled bits between many subsystems of a $N$-qubit cluster state. This makes cluster states a useful resource for teleportation and state sharing of an arbitrary two qubit state $|\psi_{2}\rangle$.  The six qubit cluster state,
\begin{equation}
      |C_6\rangle= \frac{1}{2}(|000000\rangle + |000111\rangle + |111000\rangle - |111111\rangle),
\end{equation}
has been created in laboratory conditions \cite{cluster1}. It has been shown that the four and five qubit cluster states are important resources for teleportation and QIS of an arbitrary two qubit state \cite{cluster2}. This gives us motivation to study the usefulness of $|C_6\rangle$ for several quantum protocols like teleportation, QIS, remote state preparation and dense coding. This paper is organised as follows : In the first section, we show that $|C_6\rangle$ can be used for the teleportation of an arbitrary two qubit state $|\psi_2\rangle$. It is found that $|C_6\rangle$ can teleport specific types of three qubit states e.g., GHZ states, but not an arbitrary three qubit state.  In the subsequent section, we explicate two different protocols for the QIS of $|\psi_2\rangle$ using $|C_6\rangle$ and compare their properties. We investigate
the usefulness of $|C_6\rangle$ for dense coding of classical bits. Subsequently, we show the usefulness of $|C_6\rangle$ for remote state preparation is demonstrated. 
\section{Teleportation of $|\psi_2\rangle$ using $|C_6\rangle$}
As stated earlier, $|C_6\rangle$ is an useful resource for the teleportation of $|\psi_2\rangle$, as there are
two entangled bits between several of its subsystems. In this protocol, the sender Alice possesses qubits 1, 6, 2, 5 and Bob possesses qubits 3 and 4 of $|C_6\rangle$, respectively. Alice also has the state $|\psi_2\rangle$ that
she wants to teleport to Bob. The scheme proceeds as follows: Initially, Alice performs a six qubit von-Neumann 
measurement on her qubits and then conveys her outcome to Bob via four cbits. Depending on the classical information sent by Alice, Bob can apply an appropriate unitary operator and retrieve $|\psi_2\rangle$. The outcome of the measurements performed by Alice and the state obtained by Bob are shown in Table 1. 

\begin{table}[h]
\caption{\label{tab1}Teleportation : The outcome of the measurements performed by Alice and the states obtained by Bob using the cluster state. Normalisation factors have been omitted for convenience.}
\begin{tabular}{|c||c|}
\hline {\bf Outcome of the measurement } & {\bf State obtained }\\
$[|000000\rangle + |100101\rangle + |011010\rangle + |111111\rangle]$&$\alpha|00\rangle + \mu|01\rangle + \gamma|10\rangle - \beta|11\rangle$\\
$[|000000\rangle - |100101\rangle - |011010\rangle + |111111\rangle]$&$\alpha|00\rangle - \mu|01\rangle - \gamma|10\rangle - \beta|11\rangle$\\
$[|000000\rangle - |100101\rangle + |011010\rangle - |111111\rangle]$&$\alpha|00\rangle - \mu|01\rangle + \gamma|10\rangle + \beta|11\rangle$\\
$[|000000\rangle + |100101\rangle - |011010\rangle - |111111\rangle]$&$\alpha|00\rangle + \mu|01\rangle - \gamma|10\rangle + \beta|11\rangle$\\
$[|000101\rangle + |101010\rangle + |011111\rangle + |110000\rangle]$&$\alpha|01\rangle + \mu|10\rangle - \gamma|11\rangle + \beta|00\rangle$\\
$[|000101\rangle - |101010\rangle - |011111\rangle + |110000\rangle]$&$\alpha|01\rangle - \mu|10\rangle + \gamma|11\rangle - \beta|00\rangle$\\
$[|000101\rangle - |101010\rangle + |011111\rangle - |110000\rangle]$&$\alpha|01\rangle - \mu|10\rangle - \gamma|11\rangle + \beta|00\rangle$\\
$[|000101\rangle + |101010\rangle - |011111\rangle - |110000\rangle]$&$\alpha|01\rangle + \mu|10\rangle + \gamma|11\rangle - \beta|00\rangle$\\
$[|001010\rangle + |101111\rangle + |010000\rangle + |110000\rangle]$&$\alpha|10\rangle - \mu|11\rangle + \gamma|00\rangle + \beta|01\rangle$\\
$[|001010\rangle - |101111\rangle - |010000\rangle + |110000\rangle]$&$\alpha|10\rangle + \mu|11\rangle - \gamma|00\rangle + \beta|01\rangle$\\
$[|001010\rangle - |101111\rangle + |010000\rangle - |110000\rangle]$&$\alpha|10\rangle + \mu|11\rangle + \gamma|00\rangle - \beta|01\rangle$\\
$[|001010\rangle + |101111\rangle - |010000\rangle - |110000\rangle]$&$\alpha|10\rangle - \mu|11\rangle - \gamma|00\rangle - \beta|01\rangle$\\
$[|001111\rangle + |100000\rangle + |010101\rangle + |111010\rangle]$&$-\alpha|11\rangle + \mu|00\rangle + \gamma|01\rangle + \beta|10\rangle$\\
$[|001111\rangle - |100000\rangle - |010101\rangle + |111010\rangle]$&$-\alpha|11\rangle - \mu|00\rangle - \gamma|01\rangle + \beta|10\rangle$\\
$[|001111\rangle - |100000\rangle + |010101\rangle - |111010\rangle]$&$-\alpha|11\rangle - \mu|00\rangle + \gamma|01\rangle - \beta|10\rangle$\\
$[|001111\rangle + |100000\rangle - |010101\rangle - |111010\rangle]$&$-\alpha|11\rangle + \mu|00\rangle - \gamma|01\rangle - \beta|10\rangle$\\
\hline
\end{tabular}
\end{table}
All the outcomes of the measurement are mutually orthogonal to each other indicating that
this scheme is deterministic. This successfully completes the teleportation protocol of an arbitrary $|\psi_2\rangle$ using $|C_6\rangle$.
It is worth mentioning that each of the six qubit measurement outcomes can be further decomposed into
Bell state measurements. For instance, the first measurement outcome could be further decomposed as 
\begin{eqnarray}
(|\psi_{+}\rangle|+\rangle   +   |\psi_{-}\rangle|-\rangle)(|\psi_{+}\rangle|+\rangle   +   |\psi_{-}\rangle|-\rangle) +  
(|\psi_{+}\rangle|-\rangle   +   |\psi_{-}\rangle|+\rangle)(|\psi_{+}\rangle|-\rangle   +   |\psi_{-}\rangle|+\rangle) + \\ \nonumber
(|\phi_{+}\rangle|+\rangle   -   |\phi_{-}\rangle|-\rangle)(|\phi_{+}\rangle|+\rangle   +   |\phi_{-}\rangle|-\rangle) + 
(|\phi_{+}\rangle|-\rangle   -   |\phi_{-}\rangle|+\rangle)(|-\phi_{+}\rangle|-\rangle   -   |\phi_{-}\rangle|+\rangle),
\end{eqnarray}
where $|\pm\rangle = \frac{1}{\sqrt{2}} (|0\rangle \pm |1\rangle)$. 
Hence, this protocol is experimentally feasible.

\section{QIS of $|\psi_2\rangle$ using $|C_6\rangle$}
Quantum secret sharing provides an useful tool for the sharing of both classical and quantum information using entanglement as a resource. Sharing of quantum information among a group of participants such that the original information cannot be completely
reconstructed by any one of the parties by themselves is referred to as ``quantum information splitting.'' One can devise different 
protocols for splitting of a state using the same entangled channel by redistributing the qubits among the participants.
In general, it has been proven  that one can devise $(N-2n)$ protocols for the splitting of an arbitrary $n$ qubit state among two parties \cite{sidd}.
From this theorem, we can see that one can devise two protocols for the QIS of $|\psi_2\rangle$ among two parties.

\subsection {Protocol 1}
In this protocol, Alice, possesses qubits 1, 3 Bob possesses qubits 5, 6 and Charlie possesses qubits 2, 4 in $|C_6\rangle$, respectively.
Alice possesses $|\psi_2\rangle$, that she wants Bob and Charlie to share. To achieve this purpose, Alice performs a von-Neumann, joint four partite measurement on her qubits and conveys its outcome to Charlie via four classical bits. The outcome of the
measurement performed by Alice and the corresponding Bob-Charlie system is shown in Table 2.
\begin{table}[h]
\caption{\label{tab1} Protocol I: The outcome of the measurements performed by Alice and the states obtained by Bob and Charlie using the Cluster state. Normalisation factors have been omitted for convenience.}
\begin{tabular}{|c||c|}
\hline {\bf Outcome of the measurement } & {\bf State obtained }\\
$[|0000\rangle + |1001\rangle + |0110\rangle + |1111\rangle]$&$\alpha|0000\rangle + \mu|0101\rangle + \gamma|1010\rangle - \beta|1111\rangle$\\
$[|0000\rangle - |1001\rangle - |0110\rangle + |1111\rangle]$&$\alpha|0000\rangle - \mu|0101\rangle - \gamma|1010\rangle - \beta|1111\rangle$\\
$[|0000\rangle - |1001\rangle + |0110\rangle - |1111\rangle]$&$\alpha|0000\rangle - \mu|0101\rangle + \gamma|1010\rangle + \beta|1111\rangle$\\
$[|0000\rangle + |1001\rangle - |0110\rangle - |1111\rangle]$&$\alpha|0000\rangle + \mu|0101\rangle - \gamma|1010\rangle + \beta|1111\rangle$\\
$[|0001\rangle + |1010\rangle + |0111\rangle + |1100\rangle]$&$\alpha|0101\rangle + \mu|1010\rangle - \gamma|1111\rangle + \beta|0000\rangle$\\
$[|0001\rangle - |1010\rangle - |0111\rangle + |1100\rangle]$&$\alpha|0101\rangle - \mu|1010\rangle + \gamma|1111\rangle - \beta|0000\rangle$\\
$[|0001\rangle - |1010\rangle + |0111\rangle - |1100\rangle]$&$\alpha|0101\rangle - \mu|1010\rangle - \gamma|1111\rangle - \beta|0000\rangle$\\
$[|0001\rangle + |1010\rangle - |0111\rangle - |1100\rangle]$&$\alpha|0101\rangle + \mu|1010\rangle + \gamma|1111\rangle - \beta|0000\rangle$\\
$[|0010\rangle + |1011\rangle + |0100\rangle + |1100\rangle]$&$\alpha|1010\rangle - \mu|1111\rangle + \gamma|0000\rangle + \beta|0001\rangle$\\
$[|0010\rangle - |1011\rangle - |0100\rangle + |1100\rangle]$&$\alpha|1010\rangle + \mu|1111\rangle - \gamma|0000\rangle + \beta|0001\rangle$\\
$[|0010\rangle - |1011\rangle + |0100\rangle - |1100\rangle]$&$\alpha|1010\rangle + \mu|1111\rangle + \gamma|0000\rangle - \beta|0001\rangle$\\
$[|0010\rangle + |1011\rangle - |0100\rangle - |1100\rangle]$&$\alpha|1010\rangle - \mu|1111\rangle - \gamma|0000\rangle - \beta|0001\rangle$\\
$[|0011\rangle + |1000\rangle + |0101\rangle + |1110\rangle]$&$-\alpha|1111\rangle + \mu|0000\rangle + \gamma|0101\rangle + \beta|1010\rangle$\\
$[|0011\rangle - |1000\rangle - |0101\rangle + |1110\rangle]$&$-\alpha|1111\rangle - \mu|0000\rangle - \gamma|0101\rangle + \beta|1010\rangle$\\
$[|0011\rangle - |1000\rangle + |0101\rangle - |1110\rangle]$&$-\alpha|1111\rangle - \mu|0000\rangle + \gamma|0101\rangle - \beta|1010\rangle$\\
$[|0011\rangle + |1000\rangle - |0101\rangle - |1110\rangle]$&$-\alpha|1111\rangle + \mu|0000\rangle - \gamma|0101\rangle - \beta|1010\rangle$\\
\hline
\end{tabular}
\end{table}
\\

Bob now performs a two qubit measurement on qubits 5 and 6
and conveys its outcome to Charlie via two classical bits. Depending on the outcomes of both their measurements, 
Charlie can apply an appropriate unitary transformation on his qubits to reconstruct  $|\psi_2\rangle$. 
For instance, had the Bob-Charlie system evolved into the first state given in Table 2, then
the outcome of the measurement performed by Bob and the corresponding state obtained by Charlie
is shown in the Table 3.
\begin{table}[h]
\caption{\label{tab2} State Sharing : QIS between Bob and Charlie. Normalisation factors have been omitted for convenience.}
\begin{tabular}{|c||c|}
\hline {\bf Outcome of the measurement } & {\bf State obtained }\\
$[|00\rangle + |01\rangle + |10\rangle + |11\rangle]$&$[\alpha|00\rangle + \mu|01\rangle + \gamma|10\rangle - \beta|11\rangle]$\\
$[|00\rangle + |01\rangle - |10\rangle - |11\rangle]$&$[\alpha|00\rangle + \mu|01\rangle - \gamma|10\rangle + \beta|11\rangle]$\\
$[|00\rangle - |01\rangle - |10\rangle + |11\rangle]$&$[\alpha|00\rangle - \mu|01\rangle - \gamma|10\rangle - \beta|11\rangle]$\\
$[|00\rangle - |01\rangle + |10\rangle - |11\rangle]$&$[\alpha|00\rangle - \mu|01\rangle + \gamma|10\rangle + \beta|11\rangle]$\\
\hline
\end{tabular} 
\end{table}
\\
It can be noticed that all the basis states for Alice's measurement can be further decomposed into
individual Bell basis measurements. For instance, the first measurement outcome can be further
decomposed as  $\frac{1}{2}$ $(|\psi_{+}\rangle |\psi_{-}\rangle + |\phi_{+}\rangle |\phi_{-}\rangle)$.
Since this protocol involves only Bell basis measurements, it can be experimentally realized in various systems.
We now devise another protocol for the QIS of $|\psi_2\rangle$ using $|C_6\rangle$.
\subsection{Protocol 2}
In this protocol Alice possesses qubits 1, 3 and 5 Bob possesses qubit 6 and Charlie possesses qubits 2 and 4 in $|C_6\rangle$ respectively. 
Alice performs a five partite von-Neumann measurement on his qubits and sends its qubit to Bob via four classical bits.
The outcome of the measurement performed by Alice and the corresponding Bob-Charlie system is shown in 
Table 4.
\begin{table}[h]
\caption{\label{tab1} State Sharing : The outcome of the measurements performed by Alice and the states obtained by Bob and Charlie using the cluster state. Normalisation factors have been omitted for convenience.}
\begin{tabular}{|c||c|}
\hline {\bf Outcome of the measurement } & {\bf State obtained }\\
$[|00000\rangle + |01101\rangle + |10010\rangle + |11111\rangle]$&$\alpha|000\rangle + \mu|010\rangle + \gamma|101\rangle - \beta|111\rangle$\\
$[|00000\rangle + |01101\rangle - |10010\rangle - |11111\rangle]$&$\alpha|000\rangle + \mu|010\rangle - \gamma|101\rangle + \beta|111\rangle$\\
$[|00000\rangle - |01101\rangle - |10010\rangle + |11111\rangle]$&$\alpha|000\rangle - \mu|010\rangle - \gamma|101\rangle - \beta|111\rangle$\\
$[|00000\rangle - |01101\rangle + |10010\rangle - |11111\rangle]$&$\alpha|000\rangle - \mu|010\rangle + \gamma|101\rangle + \beta|111\rangle$\\
$[|00101\rangle + |01010\rangle + |10111\rangle + |11000\rangle]$&$\alpha|010\rangle + \mu|101\rangle - \gamma|111\rangle + \beta|000\rangle$\\
$[|00101\rangle + |01010\rangle - |10111\rangle - |11000\rangle]$&$\alpha|010\rangle + \mu|101\rangle + \gamma|111\rangle - \beta|000\rangle$\\
$[|00101\rangle - |01010\rangle - |10111\rangle + |11000\rangle]$&$\alpha|010\rangle - \mu|101\rangle + \gamma|111\rangle + \beta|000\rangle$\\
$[|00101\rangle - |01010\rangle + |10111\rangle - |11000\rangle]$&$\alpha|010\rangle - \mu|101\rangle - \gamma|111\rangle - \beta|000\rangle$\\

$[|00010\rangle + |01111\rangle + |10000\rangle + |11101\rangle]$&$\alpha|101\rangle - \mu|111\rangle + \gamma|000\rangle + \beta|010\rangle$\\
$[|00010\rangle + |01111\rangle - |10000\rangle - |11101\rangle]$&$\alpha|101\rangle - \mu|111\rangle - \gamma|000\rangle - \beta|010\rangle$\\
$[|00010\rangle - |01111\rangle - |10000\rangle + |11101\rangle]$&$\alpha|101\rangle + \mu|111\rangle - \gamma|000\rangle + \beta|010\rangle$\\
$[|00010\rangle - |01111\rangle + |10000\rangle - |11101\rangle]$&$\alpha|101\rangle + \mu|111\rangle + \gamma|000\rangle - \beta|010\rangle$\\

$[|00111\rangle + |01000\rangle + |10101\rangle + |11000\rangle]$&$-\alpha|111\rangle + \mu|000\rangle + \gamma|010\rangle + \beta|000\rangle$\\
$[|00111\rangle + |01000\rangle - |10101\rangle - |11000\rangle]$&$-\alpha|111\rangle + \mu|000\rangle - \gamma|010\rangle - \beta|000\rangle$\\
$[|00111\rangle - |01000\rangle - |10101\rangle + |11000\rangle]$&$-\alpha|111\rangle - \mu|000\rangle - \gamma|010\rangle + \beta|000\rangle$\\
$[|00111\rangle - |01000\rangle + |10101\rangle - |11000\rangle]$&$-\alpha|111\rangle - \mu|000\rangle + \gamma|010\rangle - \beta|000\rangle$\\
\hline
\end{tabular}
\end{table}

Now Alice sends the outcome of her measurement using four c-bits to Charlie. Bob then performs a Hadamard measurement on his qubits
and sends his outcome to Charlie via one classical bit. For instance, had the Bob-Charlie system evolved into 
the first state shown in the Table 4, then the outcome of Bob's measurement and the corresponding state
obtained by Charlie is shown in Table 5.

\begin{table}
\caption{\label{tab5}Protocol 2. Normalisation factors have been omitted for convenience.}
\begin{tabular}{|c||c|}
\hline {\bf The outcome of Bob's measurement } & {\bf Charlie's state }\\
$(|0\rangle + |1\rangle)$&$(\alpha|00\rangle + \mu|01\rangle + \gamma|10\rangle - \beta|11\rangle)$\\
$(|0\rangle - |1\rangle)$&$(\alpha|00\rangle + \mu|01\rangle - \gamma|10\rangle + \beta|11\rangle)$\\
\hline
\end{tabular} 
\end{table}
Bob can now, perform an appropriate unitary operation and reconstruct $|\psi_2\rangle$.
Here, each measurement outcome of Alice can be further decomposed into three qubit measurements as
\begin{eqnarray}
\frac{1}{2}((|000\rangle + |111\rangle)(|00\rangle + |11\rangle)+
(|000\rangle - |111\rangle)(|00\rangle - |11\rangle) +\\\nonumber
(|011\rangle + |100\rangle)(|01\rangle + |10\rangle)+
(|011\rangle - |100\rangle)(|01\rangle - |10\rangle)).
\end{eqnarray}
As the individual GHZ measurements can be further broken down into Bell state measurements, this
scheme is also completely feasible.
\subsection{Comparison}
Its worth noting here, that the classical information to be sent to Charlie to reconstruct the state varies in the two protocols. The first measurement basis has 16 four qubit orthogonal states, whereas the second protocol uses 16 states, having 5 qubits. In the former case, the respective subsystems of Bob-Charlie composite system is more mixed as compared to the latter. Hence, we observe that Charlie has a higher probability of guessing the state without receiving Bob's classical information in the second protocol. Therefore, the first protocol is more advantageous, though it requires more classical resource. One observes a trade-off between the net classical information resource and the security of the given protocols. 

\section{Dense coding}
Dense coding is a technique of encoding classical information into quantum bits by using appropriate local
unitary operations. In general, for a given quantum state, the amount of classical bits that can be encoded
into a given quantum state $\rho^{AB}$, shared by Alice and Bob, is given by \cite{Bruss},
\begin{equation}
X(\rho^{AB})=\mathrm{log_{2}}d_{A}+S(\rho^{B})-S(\rho^{AB})
\end{equation}
Here, $d_A$ refers to the dimension of the Alice's system. By distributing the first, sixth and the fourth qubits to Alice
and the rest to Bob, we obtain the dense coding capacity of $|C_6\rangle$ to be $X(\rho^{AB}) = 3+2-0 = 5$. Hence,
Alice can send five classical bits by sending only three qubits to Bob. As in the standard dense coding protocol, Alice can encode her 
classical bits by using Pauli operators as,
\begin{equation}
U_1\otimes U_2\otimes U_3\otimes I\otimes I\otimes I\rightarrow|C_{6}\rangle_{i},
\end{equation}
where, $U_{1}$, $U_{2}$ $\in$  $\{I$, $\sigma_{1}$, $\sigma_{2}$, $\sigma_{3} \}$  and $U_{3}$ $\in$  $\{ I$, $\sigma_{1} \}$ 
and send it to Bob. Bob can either perform a cluster basis measurement or a non-destructive measurement
on the cluster state \cite{sakshi, manu, srik} and construct the classical information sent by Alice. This completes the dense coding
protocol using $|C_6\rangle$. 

\section{Remote state preparation}
Remote state preparation \cite{holypaper} refers to the teleportation of a quantum state where the state is initially known to the sender. It was shown that \cite{pati}, if the initial state to be teleported, is chosen from the real or equatorial part of the Bloch sphere, then this task can be achieved by using just one classical bit as against two classical bits, when the state is unknown to the sender. It was shown recently \cite{hello}, that remote state preparation can also be achieved using quantum information splitting. In this section, it is demonstrated that $|C_6\rangle$ is an important resource for remote state preparation. We show that if $\alpha = \beta$=$ \frac{1}{2}$ and $\mu =  \gamma = \frac{1}{2}e^{i \phi}$, then RSP can be achieved by using just two classical bits, as against four classical bits in the teleportation protocol described above. Hence, if the initial state belongs to the above states, then the classical information need not be wasted. The remote state preparation protocols proceed as follows: Initially Alice possesses qubits 1, 6, 2, 5 and Bob possesses qubits 3 and 4 and Alice wants to teleport a special class of two qubit state, discussed above to Bob. Now, Alice performs a four qubit measurement on her qubits and conveys its outcome to Bob via two classical bits. The outcome of the measurement performed by Alice and the corresponding state obtained by Bob are shown in Table 6.
\begin{table}
\caption{\label{tab6}Remote state preparation using $|C_6\rangle$. Normalisation factors have been omitted for convenience.}
\begin{tabular}{|c||c|}
\hline
$\alpha|0000\rangle + \mu*|0101\rangle + \gamma*|1010\rangle + \beta|1111\rangle$& $\alpha|00\rangle + \mu|01\rangle + \gamma|10\rangle - \beta|11\rangle$\\
$\alpha|0000\rangle + \mu*|0101\rangle - \gamma*|1010\rangle - \beta|1111\rangle$ & $\alpha|00\rangle + \mu|01\rangle - \gamma|10\rangle + \beta|11\rangle$\\
$\alpha|0000\rangle - \mu*|0101\rangle - \gamma*|1010\rangle + \beta|1111\rangle$ & $\alpha|00\rangle - \mu|01\rangle - \gamma|10\rangle - \beta|11\rangle$\\
$\alpha|0000\rangle - \mu*|0101\rangle + \gamma*|1010\rangle - \beta|1111\rangle$ & $\alpha|00\rangle - \mu|01\rangle + \gamma|10\rangle + \beta|11\rangle$\\
\hline
\end{tabular}
\end{table}
This completes the remote state preparation protocol using $|C_6\rangle$. We can also explicate protocols for RSP using QIS involving
$|C_6\rangle$ as an entangled channel. 
\section{Conclusion}

We explicated several quantum protocols using the experimentally 
achieved cluster state $|C_6\rangle$, involving largest number of entangled photons
as a quantum channel. After demonstrating that it can be used for the teleportation
of an arbitrary two qubit state, it is shown that $|C_6\rangle$ can also be used 
for the QIS of an arbitrary two qubit state in two distinct ways. We noted that, the classical information resource is not the same for different protocols involving the same states. Further, it was found that one can send five cbits by sending three quantum bits, using $|C_6\rangle$
as an entangled resource. We then explicated the usefulness of $|C_6\rangle$ for the
remote state preparation of an arbitrary two qubit state. Since, all the measurement basis
could be broken down into respective Bell state measurements, we hope that our schemes will soon be experimentally realized.


\begin{thebibliography}{26}

\bibitem{cluster1} C. Y. Lu, X. Q. Zhou, O. Ghne, W. B. Gao, J. Zhang, Z. S. Yuan, A. Goebel, T. Yang, J. W. Pan 2007
\it{Nature} \normalfont \bf{3} \normalfont {91}

\bibitem{Bennett} C. H. Bennett, G. Brassard, C. Crepeau, R. Jozsa, A. Peres, W. K. Wootters 1993 \it {Phys. Rev. Lett.} \normalfont \bf {70} 
\normalfont {1895}
\bibitem{Hillery} M. Hillery, V. Buzek, A. Berthiaume 1999 \it {Phys. Rev. A} \normalfont \bf {59} \normalfont {1829}
\bibitem{Gott} D. Gottesman, 2000 \it {Phys. Rev. A} \normalfont \bf {61} \normalfont {042311}
\bibitem{Wiesner} C. H. Bennett and S. J. Wiesner 1992 \it {Phys. Rev. Lett.} \normalfont \bf{69} \normalfont{2881}
\bibitem{one} H. J. Briegel and R. Raussendorf 2001\it {Phys. Rev. Lett.} \normalfont \bf {86} \normalfont{910}
\bibitem{wstate} W. Dur, G. Vidal and J. I. Cirac 2000 \it {Phys. Rev. A} \normalfont \bf{62} \normalfont{062314}
\bibitem{class} F. Verstraete, J. Dehaene, B. De Moor, and H. Verschelde 2002 \it {Phys. Rev. A} \normalfont \bf {65} \normalfont{052112}
\bibitem{Brown} I. D. K. Brown, S. Stepney, A. Sudbery, and S. L. Braunstein 2005 \it {J. Phys. A} \normalfont \bf{38} \normalfont {1119}
\bibitem{Borras} A. Borras, A. R. Plastino, J. Batle, C. Zander, M. Casas, and A. Plastino 2007  \it {J. Phys. A} \normalfont \bf{40} \normalfont {13407}
\bibitem{Sreramanb} S. Muralidharan, P. K. Panigrahi 2008 \it {Phys. Rev. A} \normalfont \bf {77} \normalfont {032321}
\bibitem{Sayan} Choudhury, S. Muralidharan, P. K Panigrahi 2009 \it {J. Phys. A} \normalfont \bf {42} \normalfont {115303}
\bibitem{expt} P. Walther, K. J. Resch, T. Rudolph, E. Schenck, H. Weinfurter, V. Vedral, M. Aspelmeyer  and A. Zeilinger 2005 \it{Nature} \normalfont \bf{86} \normalfont {434}{169}

\bibitem{Graph} M. Hein, J. Eisert, and H. J. Briegel 2004  \it {Phys. Rev. Lett.} \normalfont \bf  {69} \normalfont{062311}
\bibitem{Robert} R. Raussendorf, H. J. Briegel 2001 \it {Phys. Rev. Lett.} \normalfont \bf {86} \normalfont {5188}

\bibitem{Sch} D. Schlingemann and R. F. Werner 2001 \it {Phys. Rev. A} \normalfont \bf {65} \normalfont {012308}


\bibitem{local}  M. Hein, W. Dur, and H. J. Briegel 2005 \it {Phys. Rev. A} \normalfont \bf {71} \normalfont {032350}


\bibitem{sidd} S. Muralidharan, S. Karumanchi, R. Srikanth, P. K. Panigrahi \it {under preparation} \normalfont
\bibitem{cluster2}S. Muralidharan, P. K. Panigrahi \it{Phys. Rev. A} \normalfont \bf {78} \normalfont{062333}
\bibitem{Bruss} D. Bruss, G. M. D'Ariano, M. Lewenstein, C. Macchiavello, A. Sen(De), and U. Sen 2004 \it {Phys. Rev. Lett.} \normalfont \bf{93} \normalfont{210501}
\bibitem{sakshi} S. Muralidharan, S. Jain, S. Prasath E, P. K. Panigrahi \ eprint : quantph/0906.2147
\bibitem{manu}  M. Gupta, A. Pathak, and R. Srikanth 2007 \it{Int. J. Quant. Inf.} \normalfont \bf {5} \normalfont {627}
\bibitem{srik}  M. Gupta, A. Pathak, and R. Srikanth 2006 \it{Quantum Computing Back Action} \normalfont AIP
 \normalfont \bf {864} \normalfont {197}


\bibitem{holypaper}C. H. Bennett, D. P. DiVincenzo, P. W. Shor, J. A. Smolin, B. M. Terhal, and W. K. Wootters 2001 \it {Phys. Rev. Lett.} \normalfont \bf{87} \normalfont{077902}
\bibitem{pati} A. K. Pati 2000 \it {Phys. Rev. A} \normalfont \bf{63} \normalfont{014302}
\bibitem{hello}  P. K. Panigrahi, S. Karumanchi, S. Muralidharan \ eprint : quantph/0804.4215 \it{``To be published in Pramana - Journal of Physics''} \normalfont

\end{thebibliography}
\end{document}